# Laboratory Test Bench for Research Network and Cloud Computing

**Evgeniy Pluzhnik, Evgeny Nikulchev, Simon Payain**

Moscow Technological Institute, Moscow, Russia
Email: e.pluzhnik@gmail.com, nikulchev@mail.ru, sadsema@gmail.com

## Abstract

**At present moment, there is a great interest in development of information systems operating in cloud infrastructures. Generally, many of tasks remain unresolved such as tasks of optimization of large databases in a hybrid cloud infrastructure, quality of service (QoS) at different levels of cloud services, dynamic control of distribution of cloud resources in application systems and many others. Research and development of new solutions can be limited in case of using emulators or international commercial cloud services, due to the closed architecture and limited opportunities for experimentation. Article provides answers to questions on the establishment of a pilot cloud practically "at home" with the ability to adjust the width of the emulation channel and delays in data transmission. It also describes architecture and configuration of the experimental setup. The proposed modular structure can be expanded by available computing power.**

## 1. Introduction

The use of cloud technologies is rapidly increasing; many new services applying big data are appearing; there is a considerable increase in traffic, its types and quality. This trend requires bandwidth control and the development of new principles of operation of the network. Many specialists in the field of information technologies are developing to create a next-generation network Next Generation Network [1].

There are plenty of developments of new technologies that use cloud infrastructure [2], management guidelines and protocols [3] [4]. Cloud resources are recognized as an integral part of educational information systems, which is particularly important for distance learning technologies.

To check for new theoretical developments, researchers use network software emulators such as Cisco Packet

Tracert, Dynamips, GNS3 and others, giving the ability to create various types of networks and services. Software emulators have some functional limitations: closed system, a limited range of equipment and services, the complexity of embedded systems being developed.

For experimental studies with cloud and network technology, multifunctional experimental stand has been developed that lets you use the real system, to investigate the behavior of traffic and services of real systems.

Functionally test bench can be divided into three main groups, allowing you to create multi-functional system for each test problem (**Figure 1**).

Boxing statistics and management consist of:
- system of gathering and storing;
- statistical data management system;
- generation traffic system.

Altogether, the three systems make it possible to transmit to a test stand the whole picture of the real system being under examination.

*The system of collecting statistics*. Statistics from the network equipment is done by the server Paessler Router Traffic Grapher, a full, using SNMP, Nbar. The result of gathering the statistics in the database can be displayed information:
- about the download link or the selected type of traffic;
- about the queue size and the number of dropped packets;
- about all the data packets;
- about using the computing power and memory.

Statistics from servers and virtual machines using VM Stats and deployed applications allows you to get information:
- about the allocated/in use/reserved CPU;
- about the allocated/in use/reserved RAM and physical memory;
- about the database queries, search systems, services.

The information obtained is used for the analysis of the ongoing processes, verification compare performance systems, simulation and control system operation.

The control system makes the change of parameters of the system under investigation by sending commands to equipment and making changes to configuration files, in accordance with pre-set parameters.

In order to create a sequence of works, the real system, a method of network traffic generation and queries to a certain type of databases were studied, simulating the intensity at a given interval. This method is implemented in the form of a program consisting of two modules: "Generator" and "Activator". In module "Generator" generation of the files required size, defined on the basis of the analysis of statistics of the production network. In the module "Activator" a simulation was made based on previously generated data in the specified sequence after a certain interval of time.

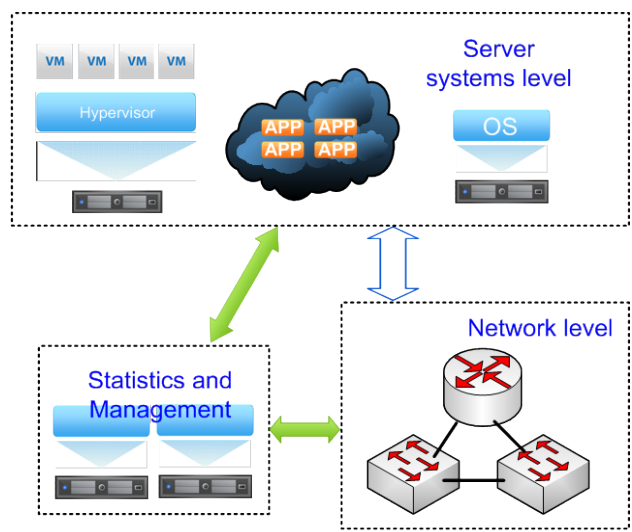

**Figure 1.** Functional diagram of the experimental laboratory bench.

## 2. Computing Architecture

A group of computing is a set of servers from different manufacturers, with different numbers of cores, memory, and storage systems allow implementing, depending on the tasks, the following functional organization.

*Server virtualization*. To conduct research on the server SunFire X2200M2 installed VMware ESXi as the place of installation used flash memory is configured virtual switch Cisco Nexus 1000 and deployed 4 virtual machines on a physical server's disks.

*Physical server*. On HP Proliant DL320 installed Microsoft Windows operating systems and collectors collection of system parameters.

*Organization of the cloud*. For the cloud is used product family VMware vCloud, allowing organizing cloud computing at all three levels. To create clouds in the experimental stand on two servers SunFire created hosts VMware ESXi system installed VCenter management, installed VMware vCloud Director, using MSSQL database to manage the cloud, as vApp applications are created virtual server with the installed applications [5].

Within each type of organization possible review new developments in the following directions:
- the analysis and management of queries and databases;
- the processing, storage and retrieval of information;
- the analysis and management of distributed computing, distribution and management of processes and requests;
- the verification and comparison of the systems under various computing resources.

Now for virtualization uses a variant of accommodation of virtual machines on a single physical server, it is planned to create the farm-assisted virtualization, storage, which will carry out experimental research related to migration of virtual machines, load balancing, and so on.

At the moment in the work of the stand involved server HP and Sun of different models. Software is used VMware vCloud, VMware vSphere, operating systems and database Microsoft, FreeBSD.

## 3. Configuring Network Data Transfer

The presence of more than 15 physical switches Cisco 29 series routers and 26 and 28 of the series, as well as virtual switches Nexus, functional network equipment, the use of dynamic routing protocols, technology Vlan, trunk, QoS and other allows to implement various schemes network for a broad range of tasks [4] [6] [7].

Existing functionality allows you to conduct research in the following directions:
- research of dynamic routing protocols;
- analysis of the traffic routing and switching;
- dynamic control downloads channels of communication, QoS priorities;
- dynamic balancing of traffic and load, a study reservation systems;
- analysis of bottlenecks and problem areas in the network part of hardware and software.

Some variants of network configurations presented in **Figure 2**, **Figure 3**.

To test the operation of storage systems and data processing [6] on the physical and virtual systems configured experimental laboratory stand (**Figure 4**).

On the server SunFire X2200M2 deployed virtualization environment VMware ESXi as the place of installation used flash memory, a pool of memory for the virtual machines are isolated from the collected RAID hard drive of the server. Created two virtual machines. VM1_db emitting 2 cores CPU and 2 GB of RAM, running Windows Server 2008R2 and database of MSSQL and VM1_file emitting 1 core CPU and 1 Gb of RAM, running Windows Server 2008R2. On a physical server HP Proliant DL320 installed operating systems Microsoft Windows Server 2008R2 and collectors collection of system parameters.

As the source host requests and files used with PCs deployed on it generating system queries and files of a given structure.

For communication with the host servers, built and configured network simulating data transmission via the Internet. Addresses configured on the router interfaces R0 and R1 and switches SW0, SW1 according to the network diagram (**Figure 3**), configured and verified routing performance of this scheme.

On a plot of the Router R1 and switch SW1 Vlan were created for each of the studied system (**Figure 5**).

All devices configured system for collecting statistics using protocol SNMP. Using switch SW_st, all devices

**Figure 2.** The organization of connection server virtual machines.

**Figure 3.** The organization of connection redundant.

**Figure 4.** Schematic experimental laboratory bench.

```
R1
        fast ethernet0/1.100
        encapsulation dot1q 100 native
        ip address
        192.168.100.1.255.255.255.0

        fast ethernet0/1.101
        encapsulation dot1q 101
        ip address
        192.168.101.1.255.255.255.0

        fast ethernet0/1.102
        encapsulation dot1q 102
        ip address
        192.168.102.1.255.255.255.0
SW1
        interface FastEthernet0/1
        switchport trunk allowed vlan
        100-102
        switchport mode trunk
```

**Figure 5.** Configuration router and switch.

connected to the server statistics collection separate channels of communication that separates the traffic from the traffic test systems for data processing and statistics. Statistic on a server operating system Microsoft Windows Server 2003R2 installed and configured service Paessler Router Traffic Grapher, providing data collection, and their representation in visual form and database.

## 4. Conclusion

Present architecture and configuration can be easily scaled to any number of resources. However, the above minimal form Laboratory Test Bench allows experiments to solve important problems related to ensuring quality of service in cloud infrastructures.

Scientific Research Publishing (SCIRP) is one of the largest Open Access journal publishers. It is currently publishing more than 200 open access, online, peer-reviewed journals covering a wide range of academic disciplines. SCIRP serves the worldwide academic communities and contributes to the progress and application of science with its publication.

Other selected journals from SCIRP are listed as below. Submit your manuscript to us via either submit@scirp.org or [Online Submission Portal](#).

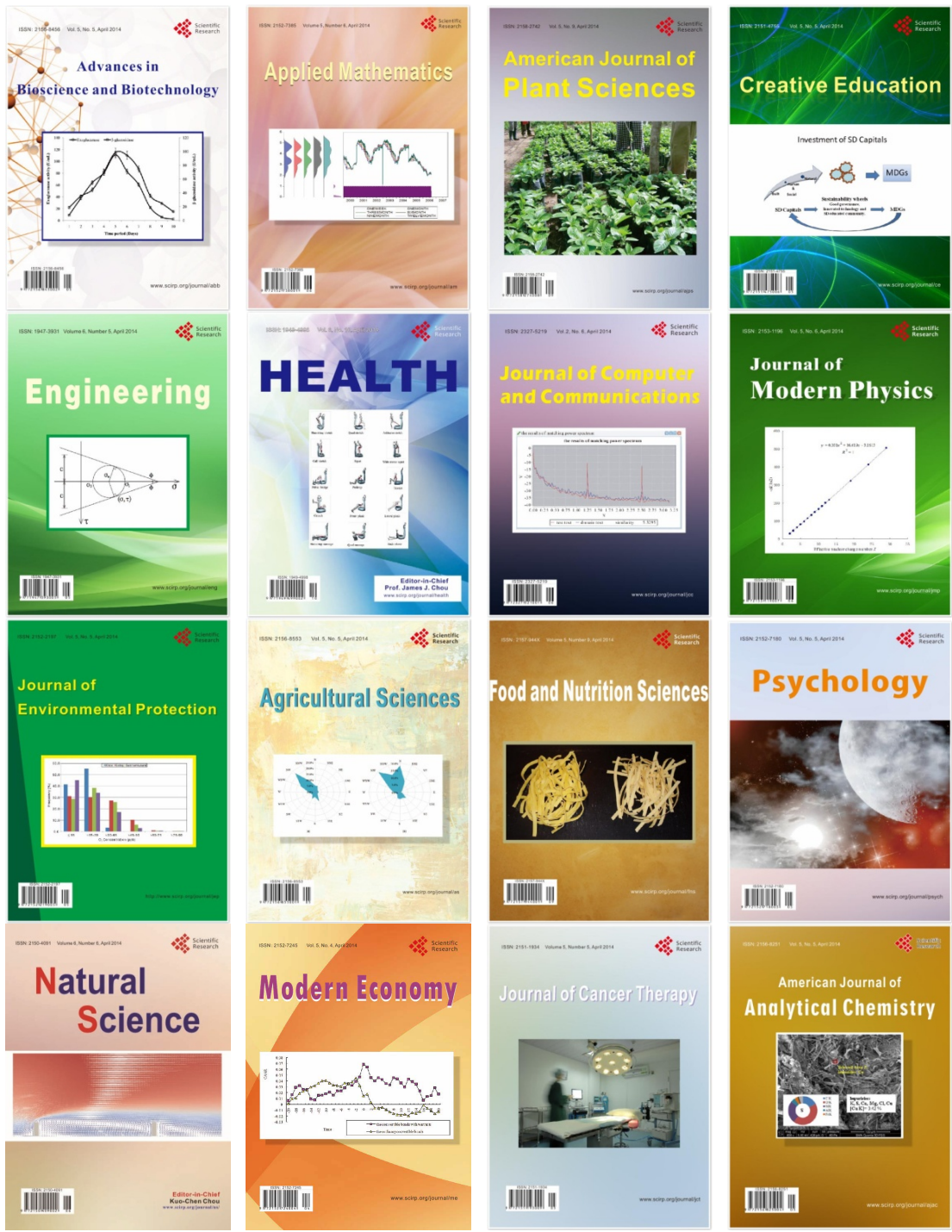